# Antinodal kink in the band dispersion of electron-doped cuprate La$_{2-x}$Ce$_x$CuO$_{4\pm\delta}$

C. Y. Tang[1,2,*], Z. F. Lin[1,2,*], J. X. Zhang[3], X. C. Guo[1,2], J. Y. Guan[1,2], S. Y. Gao[1,2], Z. C. Rao[1,2], J. Zhao[1,2], Y. B. Huang[4], T. Qian[1,5], Z. Y. Weng[3], K. Jin[1,2,5,†], Y. J. Sun[1,6,†] and H. Ding[1,2,5,†]

[1] Beijing National Laboratory for Condensed Matter Physics and Institute of Physics, Chinese Academy of Sciences, Beijing 100190, China

[2] University of Chinese Academy of Sciences, Beijing 100049, China

[3] Institute for Advanced Study, Tsinghua University, Beijing 100084, China

[4] Shanghai Advanced Research Institute, Chinese Academy of Sciences, Shanghai 201204, China

[5] Songshan Lake Materials Laboratory, Dongguan, Guangdong 523808, China

[6] Department of Physics, Southern University of Science and Technology, Shenzhen 518055, China

[*]These authors contributed equally to this work.

[†]Corresponding authors: dingh@iphy.ac.cn

**ABSTRACT**

Angle-resolved photoemission spectroscopy (ARPES) measurements have established the phenomenon of kink in band dispersion of high-$T_c$ cuprate superconductors. However, systematic studies of the kink in electron-doped cuprates are still lacking experimentally. We performed *in-situ* ARPES measurements on $La_{2-x}Ce_xCuO_{4\pm\delta}$ (LCCO) thin films over a wide electron doping ($n$) range from 0.05 to 0.23. While the nodal kink is nearly invisible, an antinodal kink around 45 meV, surviving above 200 K, is observed for $n \sim 0.05-0.19$, whose position is roughly independent of doping. The fact that the antinodal kink observed at high temperatures and in the highly overdoped region favors the phonon mechanism with contributions from the Cu-O bond-stretching mode and the out-of-plane oxygen buckling mode. Our results also suggest that the antinodal kink of LCCO is only weakly coupled to its superconductivity.

**INTRODUCTION**

The kink phenomenon in cuprates, as a sudden change in the slope of band dispersion revealed by angle-resolved photoemission spectroscopy (ARPES), remains unsolved and intriguing[1–3]. The kink is also indirectly revealed in inelastic neutron scattering (INS), inelastic x-ray scattering (IXS), Raman, and optical measurements as an anomaly in the density of states or intensities[4–7]. However, despite decades of extensive studies, the origin of kink remains controversial, electron-boson coupling (EBC), spin fluctuations, and spin-charge separation have been suggested as possible causes[2,8–10]. Among them, EBC is arguably the most preferred mechanism, with phonons, magnetic resonance modes, and polarons serving as potential candidates for the bosonic mode[11–13]. Since the electron-phonon coupling is generally regarded as the pairing glue in conventional superconductors, the bosonic mode referred from the kink in the cuprates is widely argued to play a significant role in the high-$T_c$ superconductivity.

In the most extensively studied cuprates $Bi_2Sr_2CaCu_2O_{8+\delta}$ (BSCCO) and $Bi_2Sr_{1.6}La_{0.4}CuO_{6+\delta}$, ARPES data display a strong kink with the binding energy of ~ 70 meV along the nodal direction and a weaker kink around ~ 40 meV near the antinode[5,14,15]. While the antinodal kink is widely believed to be related to the magnetic resonance mode or the out-of-plane oxygen buckling mode (O2 in the La/Ce-O plane moves parallel to the c-axis, $B_{1g}$ mode) [5,12], the nodal kink is more complicated[5,10,16,17]. In $La_{2-x}Sr_xCuO_4$, benefited from the large sample size, both ARPES and INS experiments can be performed. The nodal kink appears in LSCO at the metal-insulator transition, which is interpreted as the Cu-O bond-stretching mode (also known as the breathing mode, $E_u$ mode) or polaron mode[13,18,19]. In electron-doped cuprates $Nd_{1.85}Ce_{0.15}CuO_{4\pm\delta}$ (NCCO), $Sm_{1.85}Ce_{0.15}CuO_{4\pm\delta}$ (SCCO), and $Eu_{1.85}Ce_{0.15}CuO_{4\pm\delta}$ (ECCO), ARPES experiments suggest that the antinodal photo-hole couples to full-breathing phonons (q = [0.5, 0.5, 0], 70 meV), and the nodal photo-hole couples to half-breathing phonons (q = [0.5, 0, 0], 50 meV)[20]. The significant phonon softening along the [100] direction in many cuprates is discovered by the INS measurements[21]. Although the kink seems universal in the hole-doped and electron-doped cuprates, systematic studies of doping and temperature dependence of the kink in electron-doped cuprates are still lacking. Especially, there are no ARPES results for the kink effect in $La_{2-x}Ce_xCuO_{4\pm\delta}$ (LCCO), which has the largest rare-earth ion radius and the highest $T_c$ among the electron-doped cuprates. Motivated by this, we prepared high-quality LCCO films by the elaborate ozone/vacuum annealing method[22], and conducted ARPES measurements to systematically study the kink effect in this material.

**RESULTS**

Antinodal kink

We plot the doping evolution of the band structures of LCCO in Fig. 1. Figures. 1a, b display the band dispersions near the node and antinode with the electron doping level (*n*) varies from $n \sim 0.05$ to $n \sim 0.23$, respectively. The black lines are the band dispersions extracted from the momentum distribution curves (MDCs) by Lorentzian fitting, and the red lines are the assumed linear bare bands. The kink displays itself as a discrepancy between those two lines. It is clear that the kink is indiscernible in the nodal region, but it is ever-existing near the antinode, except for the heavily overdoped sample with $n \sim 0.23$. While in the hole-doped LSCO, the kink is more conspicuous along the nodal direction[8]. To get a better view, we show the second derivative images of the band dispersions at $n \sim 0.084$ in Fig. 1d. The dashed white lines indicate the antiferromagnetic (AF) zone boundary (AFZB), where the nodal band shows a band-folding behavior with no existence of kink. While the AF gap locates at the AFZB at the near hotspot (cut 2), the kink position at the antinode (cut 3) is away from it, which resembles the case in NCCO[23].

In general, ARPES probes the single-particle spectral function directly, and the many-body interactions exhibiting themselves as renormalization effects in band dispersion are included in the self-energy. The peak position of the real part of the self-energy (Re $\Sigma$) corresponds to the characteristic energy of the kink. We extract the antinodal band dispersions as shown in Fig. 2a and plot the Re $\Sigma$ versus $E - E_F$ in Fig. 2b, where the curves are offset vertically for clarity. It is clear that except at $n \sim 0.23$, all the samples indicate a kink at 40–50 meV, with the mean energy at $\sim 45$ meV, as summarized in Fig. 2c. In the underdoped region below 0.11, the kink shows energy softening with the increasing electron doping level, which may be related to the Coulomb screening of the increasing free charge carriers. The characteristic energy slightly increases in the overdoped region, possibly due to the less effective screening from the apical oxygens or the interaction with the O2 branch[21,24]. It is important to note that the doping evolution of the band dispersion in LCCO obeys the rigid-band shift mechanism, while in BSCCO and LSCO, the nodal dispersion is directly incompatible with that picture[10,18]. That is, in the underdoped region, the high-energy velocity increases monotonically as the hole doping level decreases and consequently surpasses the bare band velocity.

Temperature dependence of the kink

To get more information on this kink effect, we performed temperature-dependent ARPES experiments in the antinodal region of LCCO and plotted Re $\Sigma$ at different temperatures in Figs. 3a, b. The electron doping levels are estimated to be $n \sim 0.084$ and 0.175, which lies in the underdoped and the highly overdoped region of the superconducting dome of LCCO, respectively[25]. It is clear for $n \sim 0.084$ that the peak positions of Re $\Sigma$ stay at similar energies, which are represented as the red dots in Fig. 3c, and the kink survives at $T = 200$ K. The case is the same in the overdoped region at $n \sim 0.175$. For quantitative analysis, we calculate the spectral weight of Re $\Sigma$ by integrating over $E_F$ to $E_F$-200 meV and show the results in Fig. 3d. This sum area includes all the disparity between the bare band and renormalized band above -200 meV, which may be a reasonable parameter to estimate the strength of the kink. For comparison, we also extract the spectral weight of Re $\Sigma$ in the nodal region of BSCCO and LSCO[8,10]. The result suggests that the antinodal renormalization in LCCO is comparable to the renormalization effect along the nodal direction in the hole-doped cuprates. The magnitude of Re $\Sigma$ decreases slowly with increasing temperature, and its energy shows little doping dependence, resembling the behavior in the hole-doped cuprates[11,26,27].

Comparison of the kinks for different cuprates

To advance our understanding of the antinodal kink in LCCO, we plot and compare the energy scales of the kinks in different cuprate systems in Fig. 4. In the hole-doped cuprate BSCCO, the kink changes its energy from ~ 40 meV near the antinodal region to ~ 70 meV along the nodal direction[14]. While the antinodal kink is not prominent, combined ARPES and INS data have shown a ~ 70 meV in the nodal region of LSCO[4,18]. In electron-doped cuprates NCCO, SCCO, and ECCO, the 70-meV kink near the antinode is regarded as the Cu-O bond full-breathing mode[20]. Along the nodal direction, the 50-meV kink detected in NCCO and SCCO is related to the Cu-O bond half-breathing mode. In LCCO with the largest $La^{3+}$ ion radius among the lanthanide elements, besides the disappearance of the nodal kink, the antinodal kink alters to ~ 45 meV, which is downshifted from the antinodal kink energy in other electron-doped cuprates at ~ 70 meV.

**DISCUSSION**

We now try to understand the origin of this universal kink in the antinodal region in LCCO. There are several candidates as the possible origin. First of all, the widely debated magnetic resonance mode in the hole-doped cuprates is unlikely to explain the antinodal kink in LCCO. While the magnetic resonance mode in the hole-doped cuprates is around 40 meV[12,28], it is found to be ~ 10 meV in the electron-doped cuprates by neutron scattering[29,30], and this energy scale is much smaller than the observed kink energy at ~ 45 meV.

Second, some previous work pointed out that the antinodal kink in the electron-doped cuprates may originate from the opening of the AF gap[3,31,32]. We rule out this assumption for the following reasons: (1) the momentum position where the renormalization happens is away from the AFZB, as seen in Fig. 1c; (2) in the overdoped region $n \sim 0.14–0.19$ where the AF fluctuations have vanished, the kink effect remains strong. And this kink is robust at $n \sim 0.15$ at 200 K, suggesting that the proximity to the AF gap is not the mechanism.

Next, we turn to the extensively studied phonon modes. By comparing the energy scales, we found that the typical energy of the $B_{1g}$ mode is at ~ 40 meV and the Cu-O bond-stretching mode is at ~ 40–70 meV, which are close to the kink we observed at ~ 45 meV. The 70-meV kink near the antinode of NCCO, SCCO, and ECCO revealed by ARPES and INS experiments has been assigned to the oxygen full-breathing mode. Since the breathing mode is infrared-active, an optical measurement in $Pr_{1.85}Ce_{0.15}CuO_{4\pm\delta}$ (PCCO) with the $E_u$ peak at ~ 50 meV supports the mechanism of the Cu-O bond-stretching mode[33]. Besides, the $B_{1g}$ mode is Raman-active and has been confirmed in PCCO, NCCO, SCCO, and $Gd_{1.85}Ce_{0.15}CuO_{4\pm\delta}$ (GCCO) by Raman experiments[6]. A mode at 308.5 cm$^{-1}$ (~38 meV) observed in LCCO is likely ascribed to the $B_{1g}$ mode[34]. Moreover, the mode energy decreases continuously with increasing rare-earth radius, which is consistent with our data. All these results imply that the kink observed in the antinodal region in LCCO may be attributed to the $B_{1g}$ and $E_u$ phonon modes.

However, there are still some concerns. Firstly, we suggest a phonon origin based on the features of the kink, for its existence in the highly overdoped region and high temperatures. But the specific phonon mode is an assumption considering the similar energy scales. More experiments like Raman, optical, and INS in LCCO are needed to provide confirming information, though the last one may be difficult to perform due to the thin-film form. Next, the antinodal kink remains remarkably strong when superconductivity has long since faded. It seems that the antinodal renormalization is only weakly related to the high-$T_c$ superconductivity.

Moreover, the kink along the nodal direction is inconspicuous in LCCO and not prominent in NCCO/SCCO compared to that near the antinode. This behavior has a huge difference to the hole-doped cuprates, where the nodal kink is strong and holds some unique characteristics. Recent works have proposed a novel idea by assigning the anomalously nodal kink in BSCCO to the spin-charge separation[10,35]. It was suggested that quasiparticle excitations maintain coherence within the Fermi pocket, and electrons with higher binding energies than the energy of the kink decay into spinons and holons, causing the abrupt change of the bandwidth. At the antinodal region of LCCO, the decay of the quasiparticles is likely processed by phonon emission that causes the line broadening, while the nodal band dispersion remains sharp. Since the antinodal region seems conventional in LCCO, the nodal region may embody intrinsic physics, like the high-$T_c$ superconductivity and the electron-hole symmetry/asymmetry. Is there any relation between the dichotomy of the nodal and antinodal region in the hole- and electron-doped cuprates? Is the kink influenced by the distance to the cross point of AFZB and Fermi surface? What mechanisms may be concealed in the linear-like nodal band dispersion and the much smaller superconducting gap in LCCO? To answer these questions, further studies are needed.

In summary, we have observed a universal kink at ~ 45 meV in the antinodal region of LCCO, the energy scale of which is roughly doping-independent and temperature-independent. We attribute the kink to the Cu-O bond full-breathing mode and the out-of-plane oxygen buckling mode by comparing the kink energy and considering the results in other electron-doped cuprates. It implies that phonon modes are universal in both hole- and electron-doped cuprates.

The antinodal kink is also robust in the heavily overdoped region at high temperatures, indicating that it is only weakly coupled to the high-$T_c$ superconductivity.

## METHODS

High-quality LCCO thin films were grown by pulsed laser deposition (PLD) with Ce concentrations $x$ = 0.1, 0.15, and 0.19. The electron doping level, determined by oxygen and Ce stoichiometry, is estimated by fitting the Fermi surface volume using the tight-binding model. By the two-step ozone/vacuum annealing method described previously, we were able to measure ARPES spectra on LCCO films from $n \sim$ 0.05 to 0.23[22]. For $n \sim$ 0.05–0.19, we performed ARPES measurements with a Scienta R4000 analyzer and a VUV helium light source. He-IIα resonant line (40.8 eV) was used and the base pressure of the ARPES system is at $\sim 4 \times 10^{-11}$ torr. To ensure that the samples at all doping levels are in the normal state, all measurements were carried out at 30 K, unless otherwise specified. For $n \sim$ 0.23, ARPES spectrum was measured at the Dreamline beamline of the Shanghai Synchrotron Radiation Facility (SSRF) with a Scienta Omicron DA30L analyzer at $\sim$ 20 K, with the photon energy of 55 eV.

## DATA AVAILABILITY

Data are available from the corresponding author upon reasonable request.

**ACKNOWLEDEMENTS**

We thank K. Jiang and Y. G. Zhong for valuable discussions. This work was supported by the grants from the Natural Science Foundation of China (11888101, 11227903 and U1875192), the Ministry of Science and Technology of China (2016YFA0401000, 2017YFA0302902 and 2017YFA0403401), the Chinese Academy of Sciences (XDB07000000, XDB25000000).

**AUTHER CONTRIBUTIONS**

C. Y. T., Y.-J. S. and H. D. designed the experiments and supervised the project. Z. F. L. and K. J. grew the thin films, C. Y. T. performed ARPES measurements with the assistance of X. C. G., S. Y. G., Z. C. R., J. Z., Y. B. H. and T. Q. C. Y. T. analyzed the ARPES data. J. X. Z. and Z. Y. W. provided the theoretical advices. C. Y. T., J. Y. G. and H. D. wrote the manuscripts with help from all the other authors.

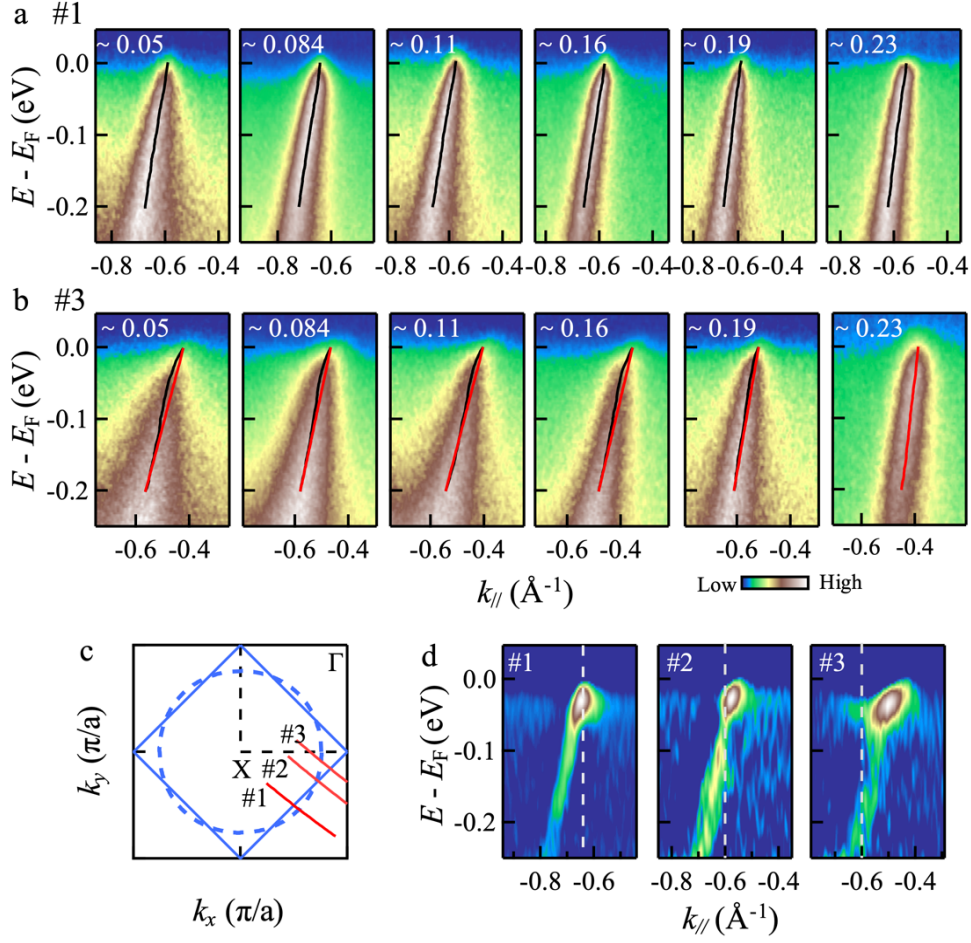

Fig. 1 **Band dispersions in LCCO. a** Nodal band dispersion of LCCO for $n \sim 0.05-0.23$. The momentum position of these bands is indicated as cut 1 in **c**. ARPES spectrum at the node of $n \sim 0.084$ was performed at 10 K. **b** Band dispersion near the antinode, represented as cut 3 in **c**. For $n \sim 0.05-0.11$, samples were ozone/vacuum annealed from as-grown films with Ce concentration $x = 0.1$, and for $n \sim 0.14-0.19$, samples were annealed from one film with $x = 0.19$. The band structures at $n \sim 0.23$ are from our previous work[22]. The black lines are the peaks of the MDCs extracted by Lorentzian fitting, and the red lines are the assumed linear bare bands. **c** Schematic of the Fermi surface at $n \sim 0.084$. Cut 1, cut 2, and cut 3 indicate the momentum cuts along the nodal direction, the near hotspot, and the antinode. **d** Second derivative images of the band dispersions at $n \sim 0.084$ as marked in **c**. The white dashed lines indicate the antiferromagnetic zone boundary.

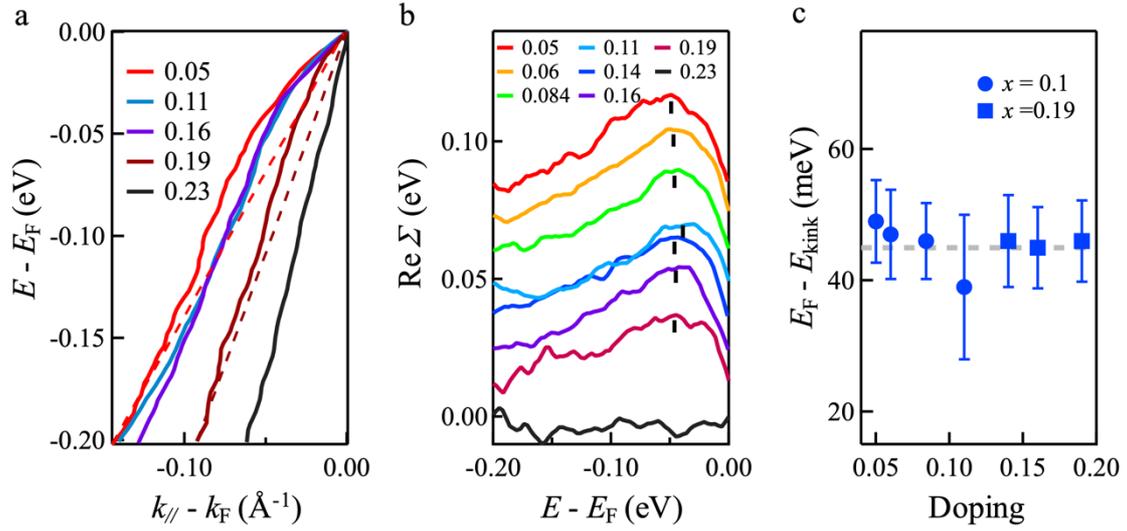

Fig. 2 **Universal antinodal kink in LCCO. a** MDC-derived antinodal band dispersions at different doping levels are displayed as the solid lines, and the dashed lines represent the linear bare bands connecting the $E_F$ to $E_F$ - 200 meV. **b** Calculated real part of the self-energies, the curves are offset vertically for clarity. **c** Doping dependence of the peak position of Re $\Sigma$, the mean energy of the kink is at ~ 45 meV. Error bars are the overestimated uncertainty of determining the peak positions.

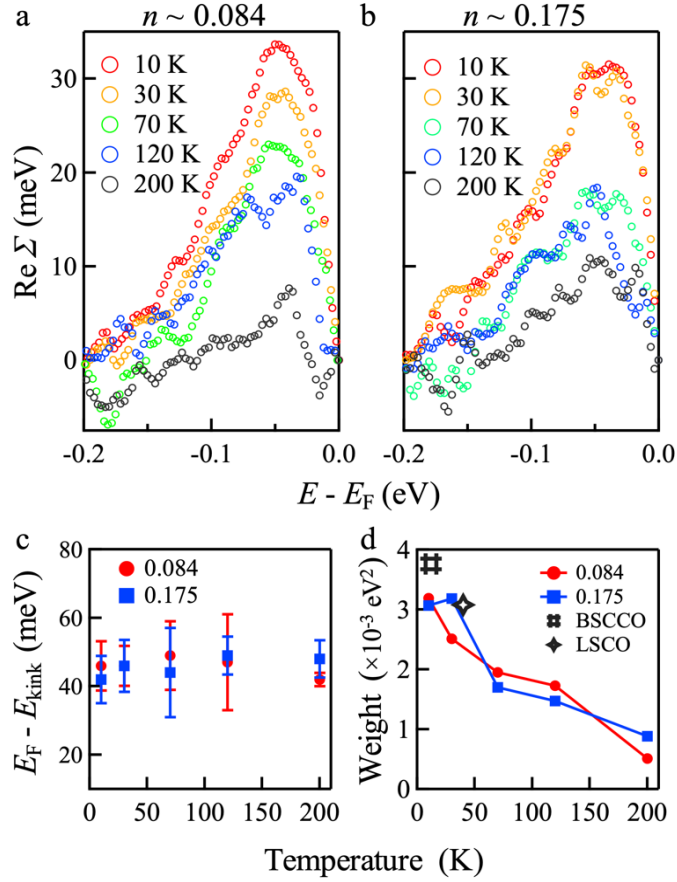

Fig. 3 **Temperature dependence of the kink. a**, **b** Temperature-dependence of the real part of the self-energy for $n \sim 0.084, 0.175$, respectively. The sample at $n \sim 0.175$ was ozone/vacuum annealed from an as-grown film with Ce concentration $x = 0.15$. **c** Extracted peak positions of the Re $\Sigma$ for $n \sim 0.084$ (red dots) and $0.175$ (blue squares) at different temperatures. **d** Spectral weight of Re $\Sigma$ by integrating over $E_F$ to $E_F - 200$ meV for $n \sim 0.084$ and $0.175$. The spectral weight of BSCCO and LSCO are presented for comparison.

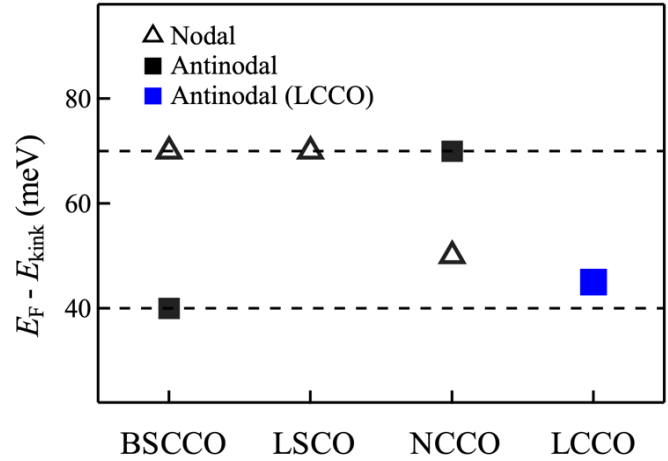

Fig. 4 **Comparison of the kink energies.** The energy scales of the kinks along the nodal direction (open triangle symbols) and near the antinodal region (solid square symbols) in the hole- and electron-doped cuprates. The dashed lines at 40 and 70 meV represent the characteristic energies of the kinks. In LCCO (indicated as the red star), the nodal kink is disappeared, but the antinode has a universal kink at ~ 45 meV.